# 3-D Lattice Simulation of the Electroweak Phase Transition at Small Higgs Mass

E.-M. Ilgenfritz[1], J. Kripfganz[2], H. Perlt[3] and A. Schiller[3]

[1] *Institut für Physik, Humboldt–Universität zu Berlin*
[2] *Institut für Theoretische Physik, Universität Heidelberg*
[3] *Fakultät für Physik und Geowissenschaften, Universität Leipzig*

June 14, 1995

**Abstract**

We study the electroweak phase transition by lattice simulations of an effective 3-dimensional theory, for a Higgs mass of about $35 GeV$. In the broken symmetry phase our results on masses and the Higgs condensate are consistent with 2-loop perturbative results. However, we find a non-perturbative lowering of the transition temperature, similar to the one previously found at $m_H = 80 GeV$. For the symmetric phase, bound state masses and the static force are determined and compared with results for pure $SU(2)$ theory.

There are strong indications that the electroweak standard theory predicts a first order phase transition at the electroweak scale [1]-[8]. The generation of the baryon asymmetry of the universe at the electroweak phase transition is an exciting hypothesis. A better quantitative characterization of the electroweak phase transition is required, however, in order to clarify whether the known baryon asymmetry could actually be generated in this way.

The electroweak phase transition cannot be treated completely by perturbative techniques. Within this frame, the symmetric phase would consist of massless W bosons. Then there are infrared problems which prevent a perturbative evaluation of the free energy to higher loop order. The true, non-perturbative behaviour of the symmetric phase will be characterized by massive $W$- and Higgs bound states instead of massless $W$ gauge bosons, however. Condensate formation may also lower the free energy of the symmetric phase. Therefore, non-perturbative techniques are required in order to determine the critical temperature at which the free energies of both phases are equal. The free energy of the broken symmetry phase can be evaluated perturbatively, as long as the Higgs mass is not too large. If Higgs particles are heavy the broken phase will also be strongly coupled, for temperatures close to the critical one. This problem will not be considered in the present paper. Our analysis is limited to a study of the phase transition for $m_H$ in the range of $35 GeV$. Our aim is to shed light on the non-perturbative behaviour of the symmetric phase, and to test the (dimensionally reduced) method through a comparison with the perturbative treatment of the broken phase where the latter should be appropriate.

The electroweak phase transition has been investigated by $3 + 1$-dimensional lattice simulations before [4, 5, 7]. Since the interesting dynamics is expected to be carried by the static (constant in $\tau$) low momentum modes, dimensional reduction should be possible, also in the neighbourhood of the phase transition. In this case lattice simulations could be done with an effective 3-dimensional action generated by integrating out the non-static modes including the fermionic ones. Other than in the case of QCD, dimensional reduction should work for the electroweak theory in the interesting temperature range because $g^2$ is small. For that reason, the non-local 3-dimensional effective action may be approximated locally, even near to the critical temperature. In connection with the electroweak phase transition this approach has been pioneered by Farakos et al. [3, 6, 8].

We study the $SU(2)$–Higgs system with one complex doublet of variable length. The gauge field is represented by the unitary $2 \times 2$ link matrices $U_{x,\mu}$ and the Higgs fields are written as $\Phi_x = \rho_x V_x$ ($\rho_x^2 = \frac{1}{2} Tr(\Phi_x^+ \Phi_x)$ is the scalar 'Higgs length', $V_x$ an element of the group $SU(2)$). The lattice action is

$$S = \beta_G \sum_p (1 - \frac{1}{2} Tr U_p) - \beta_H \sum_l \frac{1}{2} Tr(\Phi_x^+ U_{x,\mu} \Phi_{x+\mu}) + \sum_x (\rho_x^2 + \beta_R (\rho_x^2 - 1)^2) \quad (1)$$

with

$$\beta_G = \frac{4}{a g_3^2}. \quad (2)$$

In three dimensions the lattice Higgs self–coupling $\beta_R$ is given by

$$\beta_R = \frac{\lambda_3}{g_3^2} \frac{\beta_H^2}{\beta_G} . \quad (3)$$

$g_3^2$ and $\lambda_3$ denote the $3-d$ continuum gauge and Higgs self couplings which are $3-d$ renormalization group invariants. They are related to the corresponding four dimensional couplings via
$$g_3^3 = g^2 T \ , \quad \lambda_3 = T(\lambda + O(g^4)) \ . \tag{4}$$

The string operators of length $L = 1, 2, ...$ are defined as
$$E_l(L) = \Phi_x^+ U_{x,\hat{\mu}} U_{x+\hat{\mu},\hat{\mu}} U_{x+2\hat{\mu},\hat{\mu}} ... U_{x+(L-1)\hat{\mu},\hat{\mu}} \Phi_{x+L\hat{\mu}} \tag{5}$$
and used to form Higgs and $W$-operators. Actually the $3-d$ masses (inverse correlation lengths) have been obtained from the connected correlators between separated equal 'time' (in $2+1$ dimensions) planes of the sum of 'spatial' string operators within these planes projecting out the proper $SU(2)$ and spin content. To increase the overlap with the lowest mass states we choose $L = 4$ for the $W$-mass determination and also partly in the Higgs mass sector (there also $\rho_x^2$ is used). As expected, the signal is cleaner and stronger in the low temperature phase. To determine the static force we have used Wilson loops $W(R, T)$ of asymmetric extensions $5 \leq R \leq T \leq N/2$ ($N^3$ is the lattice size).

The vectorized Monte Carlo algorithm combines a three-dimensional Gaussian heat bath for the gauge fields and a four-dimensional Gaussian heat bath for the Higgs field. To reduce the autocorrelations near to the phase transition a heat bath step was followed by several reflections (eight in practice) for the Higgs and one reflection for the gauge field.

We consider the quartic coupling $\frac{\lambda_3}{g_3^2} = 0.0239$. According to the tree level based relation
$$\frac{\lambda_3}{g_3^2} \approx \frac{\lambda}{g^2} \approx \frac{1}{8} \frac{m_H^2}{m_W^2} \tag{6}$$
this would correspond to the case of zero temperature masses $m_H = 35 GeV, m_W = 80 GeV$. One should notice, however, that eq. (6) is only approximately true. It depends on details of the dimensional reduction (loop order, corrections for the adjoint Higgs field which we neglect altogether). Comparison with other calculations should therefore be done at the corresponding $\lambda$ value, and not necessarily refering to $m_H = 35 GeV$. The lattice Higgs self coupling $\beta_R$ is not fixed in this setting but runs with $\beta_H$ according to eqs. (3,6).

In the simulations we used $20^3$ and partly $30^3$ lattices. Typically 20,000 to 100,000 iterations at the smaller lattices and up to 30,000 iterations at the larger ones have been performed for every coupling pair $(\beta_H, \beta_R)$ at given $\beta_G$ (with correlator measurements every 10 iterations). Near to the phase transition the largest integrated autocorrelations

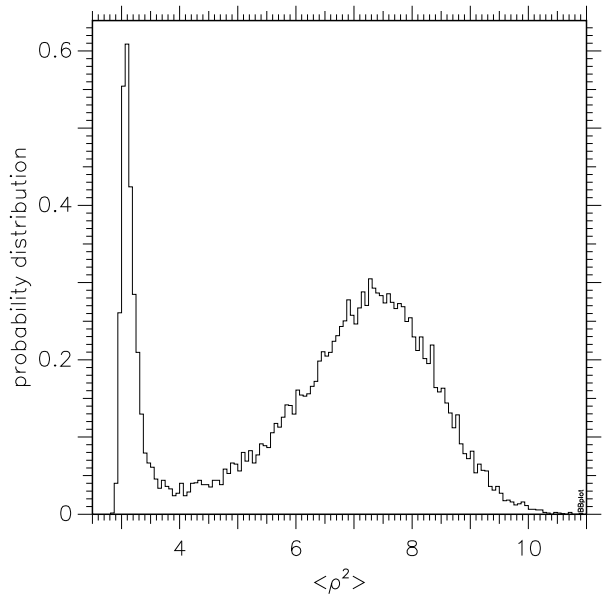

Fig.1

*Two states signal in the $<\rho^2>$ histogram on a $20^3$ lattice*

have been observed in observables containing Higgs field degrees of freedom. We were cautious to have the measured correlation lengths (up to 10) on the $20^3$ lattices not to be strongly influenced by the lattice size. We checked a few of the largest correlation lengths on larger lattices and found no significant lattice size behaviour. A particular example is presented later in Fig. 2 .

Note that we never have to specify the value of the $4-d$ gauge coupling $g^2$. Masses are obtained in units of $g^2 T$ after eliminating the lattice constant $a$ through eqs. (2,4).

We now turn to a discussion of results. Near to a first-order phase transition the two phases will coexist. If the transition is strong enough the two phases should manifest themselves as a two-state signal in the histograms of suitable observables. In Fig. 1, we show a histogram for $<\rho^2>$ (which is the length squared of the Higgs field $\rho_x^2$ averaged over the lattice for each measured configuration). For the particular lattice parameters chosen ($\beta_G = 16, \beta_H = 0.33930$) the peak of the broken phase has obviously more weight indicating that we are still slightly below the critical temperature. Locating the critical temperature precisely is a difficult problem. As it is known from previous studies [4] different indicators like the specific heat or the Binder cumulant result in various pseudo-critical couplings in a finite volume. A detailed finite size analysis is necessary to locate the inifinite volume critical parameters and to avoid these ambiguities.

From the position of the peak of the broken phase we can determine the renormalized value of the Higgs condensate, according to

$$\frac{v_R^2}{g^2 T^2} = \frac{\beta_H \beta_G}{4}(\langle \rho^2 \rangle - \langle \rho_{sym}^2 \rangle) \qquad (7)$$

where $\langle \rho_{sym}^2 \rangle$ is the expectation value of $\rho^2$ in the symmetric phase (close to the critical point) where it is actually almost independent of $\beta_H$. The relation (7) neglects a small temperature dependent two-loop contribution. In subtracting $\rho^2$ of the symmetric peak we define the condensate to be zero in the symmetric phase. This cannot be strictly maintained because $v_R$ is not renormalization group invariant and depends on the $3-d$ renormalization scale $\mu_3$ at two-loop order. The resulting condensate value is very small for any reasonable choice of $\mu_3$, however.

Equation (7) allows us to measure directly the quantity $\frac{gT}{v_R}$ which appears as coupling in quasiclassical estimates of the sphaleron rate. The sphaleron rate is sufficiently small (in order to avoid washing out any generated asymmetry) as long as $\frac{gT}{v_R}$ is of the order of $\frac{2}{3}$ (or smaller) at the critical temperature.

The best way to compare the predictions of continuum perturbation theory for the phase transition in any detail with lattice results would be to calculate the effective potential to two loop order, using both ways of regularization. This is not yet available. Therefore, we compare the relation between two observables in the two approaches. This does not require to relate explicitely the bare mass squared of the continuum formulation $m_3^2$ to the gauge-Higgs coupling $\beta_H$. We also do not have to introduce the temperature explicitely.

Fig. 2 gives a summary of our mass measurements at $\beta_G = 12$. With the operators under consideration we find two massive bound states in the symmetric phase, instead of the massless $W$ bosons one would naively expect from perturbation theory. As long as the system tunnels between the two phases a safe measurement of the Higgs correlation length is not possible while the lightest $W$-state is not influenced by this tunneling. By inspecting the Monte Carlo histories of various operators it was possible to extract subsamples to perform correlation measurements which refer to the pure phases. This was the way to clearly identify a jump of the Higgs mass between the low and high temperature phases, respectively. In the $W$-channnel we were not able to establish an analogous discontinuity within our errors. Measurements on the larger lattice ($30^3$) do not indicate severe finite size effects near to the transition. There is also no indication that a heavy state becomes much lighter going to a larger volume. This would be the way a light state would appear in a small volume expansion.

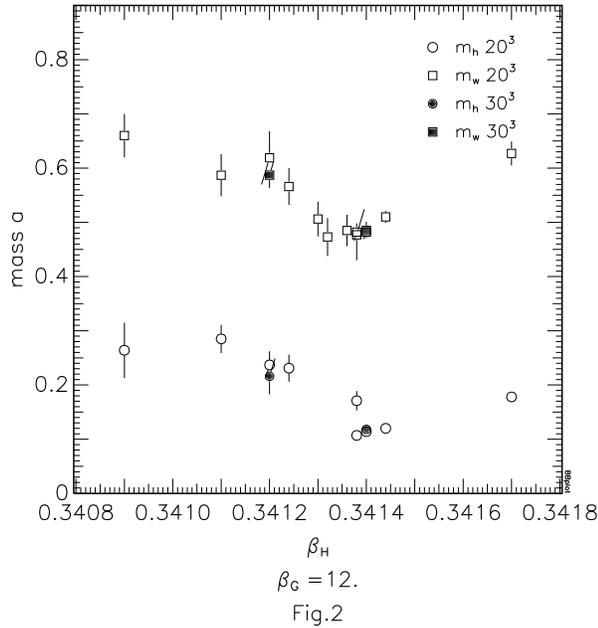

Fig.2

*Higgs and W masses in lattice units vs. gauge Higgs coupling $\beta_H$*

Table 1: Measured $W$-masses at critical $\beta_H$ vs. $\beta_G$ on a $20^3$ lattice

| $\beta_G$ | $\beta_H$ | $m_w a$ | $m_w/(g^2 T_c)$ |
|---|---|---|---|
| 12 | 0.34138(2) | 0.482(16) | 1.446(48) |
| 16 | 0.33930(4) | 0.333(17) | 1.332(68) |
| 20 | 0.33806(4) | 0.261(14) | 1.305(70) |

To identify finite $a$ effects one has to compare masses at the same temperature. To get an idea about the strength of these effects we compare measured $W$-states at couplings $\beta_H$ which are, within our precision, nearest to the critical ones and, therefore correspond to the same critical temperature. For each $\beta_G$, we select the suitable $\beta_H$ value according to the largest correlation length in the Higgs sector. The data shown in Table 1 are consistent with each other for $\beta_G = 16$ and 20 while that at $\beta_G = 12$ significant finite $a$ corrections cannot be ruled out. One should bear in mind, however, that in the present stage we cannot make sure that all these measurements correspond to exactly the same temperature with sufficient accuracy.

We now discuss the mass measurements in the broken phase. One important quantity to judge the reliability of perturbation theory is the wave function renormalization for the

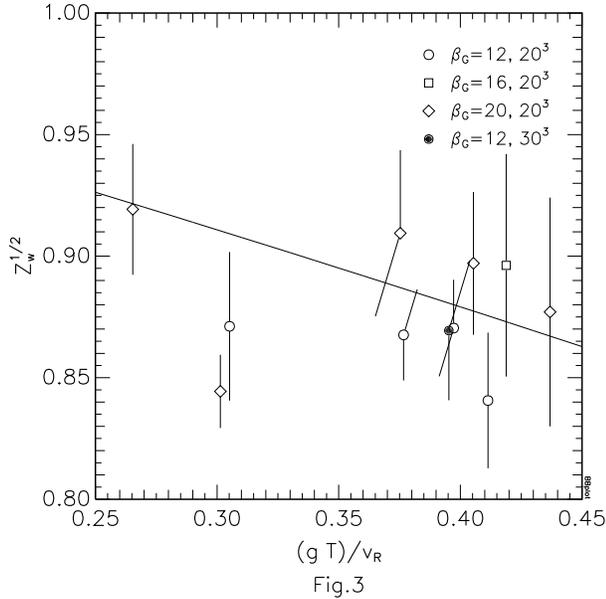

Fig.3

*Lattice results for the wave function renormalization of the W field vs. $(gT)/v_R$ compared to one-loop perturbative predictions*

W field. It can be measured by relating the W mass to the renormalized condensate $v_R$

$$m_w = Z_w^{-\frac{1}{2}} \frac{g v_R}{2}. \qquad (8)$$

Results are shown in Fig. 3, together with the one-loop prediction calculated in Feynman gauge [9]. Within the statistical errors agreement is found. $Z_w$ is close enough to one to expect that the broken phase itself can be understood perturbatively in this range of light Higgs masses. Again, the data points refering to $\beta_G = 12$ are systematically somewhat below the perturbative curve, which can be interpreted as indication for a finite $a$ correction at the strongest gauge coupling that we have investigated.

The Higgs mass $m_h$ constitutes the other piece of information we can use to test the perturbative behaviour of the broken phase. Results are shown in Fig. 4, for $\frac{m_h}{m_w}$ plotted versus the corresponding measured value of $\frac{gT}{v_R}$. Within errors, we again find agreement with perturbative results obtained from the Feynman gauge two-loop potential [10]. The perturbative Higgs mass is obtained through the second derivative of the potential at the broken minimum (i.e. at zero momentum). This might account for a slight but systematic trend of the perturbative mass ratio to be larger than the lattice data which correspond to pole masses. The perturbative masses are renormalized by the corresponding one-loop

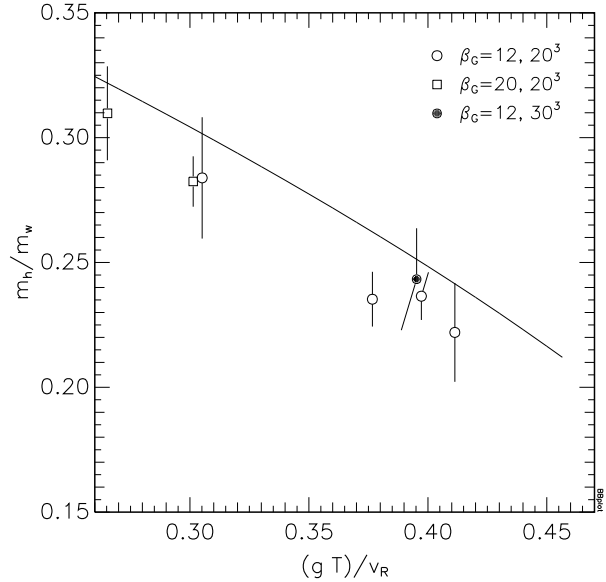

Fig.4

*Measured mass ratio as function of $(gT)/v_R$ compared to perturbation theory*

wave function renormalizations $Z_h^{-\frac{1}{2}}$ and $Z_w^{-\frac{1}{2}}$, resp., in Feynman gauge taken from refs. [9, 10].

Finally, we discuss the static force. Results are shown in Fig. 5 for two values of $\beta_H$ corresponding to the broken and symmetric phase, respectively. Using the measured Wilson loops to determine the static potential and the force from Creutz ratios would result in an overestimation of the errors. Therefore we have fitted the large 'time' dependence of prolongated Wilson loops by an exponential in order to determine the ground state energy of the pair. Using these fitted energies with their errors we have defined the static force. The measured static force obtained from the naive Creutz ratios are in good agreement with this prescription. In Fig. 5, the perturbative one-loop contribution in the broken phase

$$V_{p.t.}(R) = \frac{g_3^2 C_F}{L^2} \sum_{k^{(i)}} \frac{e^{ik_2^{(i)} R}}{m_w^2 + 4 \sum_{\mu=1}^2 \sin^2(k_\mu^{(i)}/2)} \ , \ \ C_F = \frac{3}{4} \qquad (9)$$

is also shown. The measured value of $m_w$ has been plugged into this expression. For the $3-d$ gauge coupling, two values have been used: the bare $g_3^2$ as well as an effective coupling $g_{3R}^2$ describing best the data. This effective coupling sums up finite loop contributions and is in good agreement with the reported wave function renormalization $Z_w$.

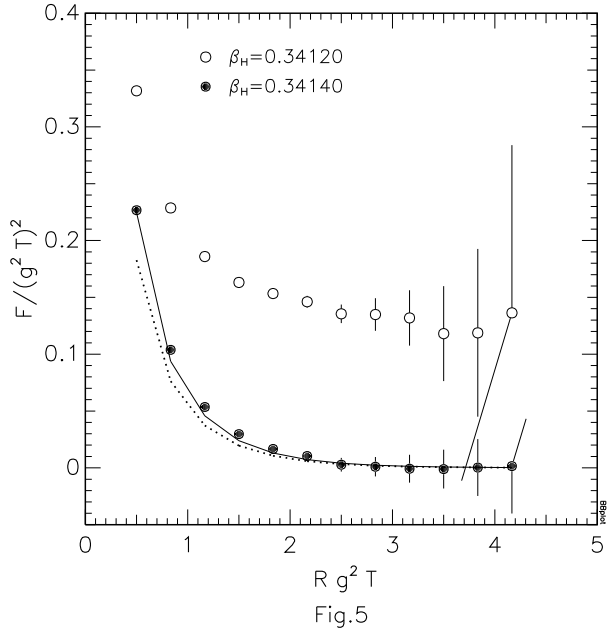

Fig.5

*Static force vs. distance in symmetric (open circles) and broken (filled circles) phase at $\beta_G = 12$. The solid (dotted) line is the perturbative one loop contribution with $g_3^2 a = 0.41(0.33)$, $m_w a = 0.485$.*

Whereas the static force is well represented perturbatively in the broken phase, a non-perturbative contribution is clearly seen in the symmetric phase. Roughly, it can be described as a linear contribution to the potential, with a string tension being in good agreement with $\sigma = 0.11(g^2T)^2$ found by Teper ($0.13(g^2T)^2$ at $\beta_G = 12$) [11] in $3-d$ pure $SU(2)$ gauge theory. This confining-type behaviour should not extend to very large distances beyond a screening length. Screening is not seen, however, up to the distances we could measure.

The remaining feature to be discussed in connection with the phase transition is an eventual lowering of the critical temperature due to nonperturbative effects (condensates) present in the symmetric phase. Returning to Fig. 4 we notice the perturbative two-loop estimate of the phase transition to occur at about $\frac{gT}{v_R} = 0.46$. This is the endpoint of the line. Lattice results point towards $\frac{gT}{v_R} = 0.41$ instead. We can now simply calculate the free energy that corresponds to that latter value to 2-loop accuracy in the broken phase. This value can then be identified with the non-perturbative (condensate) contribution to the free energy of the symmetric phase. The result is

$$\Delta F = -0.027(g^2T)^3 \qquad (10)$$

somewhat depending on the gauge (results shown represent an average of Landau and Feynman gauge). This value is very sensitive to the value of $\frac{gT}{v_R}$ considered to be the critical one. $\frac{gT}{v_R} = 0.42$ would give $\Delta F = -0.019(g^2T)^3$ instead. This should be compared with the value of $\Delta F = -0.016(g^2T)^3$ given in ref. [6] for $m_H = 80 GeV$. Thus, we confirm the presence of such a term which seems to be more or less independent of the Higgs self-coupling. Therefore, it should essentially be understood in terms of the dynamics of $3-d$ pure gauge theory alone.

In summary, our lattice results are in reasonable agreement with perturbative predictions for the Higgs phase, as they should. The symmetric phase shows a confining behaviour at the intermediate distances we have explored. The string tension is very close to the value found for the $3-d$ pure $SU(2)$ theory. Two massive bound states are identified in the symmetric phase. The mass of Higgs state is smaller than that of the lightest glueball of the pure $SU(2)$ theory. The lowering of the critical temperature, presumably induced by gauge condensates in the symmetric phase, shows little dependence on the Higgs self-coupling $\lambda$.

**Acknowledgements:** E.M. I. is supported under DFG grant Mu932/1–2, H. P. under DFG grant Il29/1–2, and J. K. under DFG grant We1056/2–3. We thank the council of HLRZ Jülich for providing CRAY-YMP resources.